\documentclass[twocolumn,showpacs,aps,prd]{revtex4}
\pdfoutput=1
\newcommand{\half}{\frac{1}{2}}
\newcommand{\thrd}{\frac{1}{3}}
\newcommand{\frth}{\frac{1}{4}}
\newcommand{\sxth}{\frac{1}{6}}
\newcommand{\Da}{\Delta a}
\newcommand{\dar}{\partial_r}
\newcommand{\R}{{\bar r}}
\newcommand{\z}{{\bar z}}
\usepackage[dvipsnames]{color}
\newcommand{\revision}[1]{{           {#1}}} 
\begin{document}
\title{Conformal theory of gravitation and cosmic expansion}
\author{R. K. Nesbet }
\affiliation{
IBM Almaden Research Center,
650 Harry Road,
San Jose, CA 95120-6099, USA
\begin{center}rkn@earthlink.net\end{center}
}
\date{\today}
\begin{abstract}
The postulate of universal Weyl conformal symmetry for all elementary
physical fields introduces nonclassical gravitational effects in both 
conformal gravitation (CG) and the conformal Higgs model (CHM).  The
resulting theory is found to explain major observed phenomena including 
excessive galactic rotation velocities and accelerating Hubble 
expansion, without invoking dark matter (DM). The recent history of this
development is surveyed here.  The argument is confined to implications
of classical field theory, which include galactic baryonic Tully-Fisher 
relations and dark galactic haloes of definite large radius. 
Cosmological CHM parameters exclude a massive Higgs boson but are 
consistent with a novel alternative particle of the observed mass.\\
Keywords: conformal gravity theory; conformal Higgs model; 
 depleted galactic halo model
\end{abstract}
\pacs{04.20.Cv,98.80.-k,98.62.Gq} 
\maketitle
\section{Introduction}
\par Conformal gravity (CG), reviewed in detail\citep{MAN06,MAN12} by
Mannheim, replaces the Einstein-Hilbert Lagrangian density by a 
quadratic contraction of the conformal Weyl tensor\citep{WEY18,WEY18a}, 
invariant under Weyl scaling by an arbitrary scalar field.  The implied 
Schwarzschild equation, in spherical geometry\citep{MAK89,MAN91}, fitted 
to observed galactic rotation\citep{MAN06}, determines parameters with 
no need for dark matter (DM).
\revision{
\par Conformal theory is limited here to symmetry defined 
by invariance of the coupled field action integral under Weyl scaling,
defined below, without altering the postulated Riemannian geometry of
Einstein-Hilbert general relativity.  The further development of
Weyl geometry\cite{JW18}, possibly relevant to the intense gravitation 
of black holes and the big bang model, is not considered here. 
}
\par The Weyl tensor vanishes identically in the isotropic uniform
geometry postulated to describe cosmic Hubble expansion, precluding
explanation by CG.  However, if the Higgs scalar field of particle
theory $\Phi$ is postulated to have Weyl scaling symmetry, its 
Lagrangian density acquires a gravitational term\citep{MAN06}. The 
resulting conformal Higgs model (CHM) implies a modified Friedmann 
equation consistent with observed Hubble expansion\citep{NESM1}. 
\par In spherical geometry, the CHM Friedmann equation determines 
luminosity distance as a function of redshift $z\leq 1$ for observed 
supernovae as accurately as any alternative theory\citep{NESM1,NES13}. 
The Higgs symmetry-breaking mechanism produces the gravitational 
effect of a dark energy source cosmological constant\citep{NESM1}.  
Centrifugal cosmic acceleration persists in present epoch $z\leq 1$.
CG and CHM together have been found to determine basic parameters of 
the Higgs model from observed galactic and cosmological data. 
These results are reviewed here. 

\section{Qualitative regularities of cosmological data}
\par Radial centripetal acceleration $a$ implies $v^2=ra$ for velocity   
of circular geodesics.  When it became possible to observe and measure 
velocities of matter in stable circular galactic orbits, $v^2$ was found 
to exceed Newtonian $ra_N$ of observed baryonic matter by quite large 
factors, and to remain nearly constant (flat) as $r$ increases. This was 
well before the observation of galactic haloes with implied fixed cutoff 
boundaries. This motivated the MOND model\citep{MIL83,FAM12}. 
MOND implies flat $v(r)$ for large $r$. 
For $a_N\ll a_0$, MOND $v^4=a^2r^2\to GMa_0$, the empirical
baryonic Tully-Fisher relation\citep{TAF77,MCG05,MCG11,OCM18}.
\par It has recently been shown\citep{MLS16} for 153 disc galaxies                                                                                                                                                                            that observed acceleration $a$ is effectively a universal function of 
Newtonian acceleration $a_N$, computed for the observed baryonic  
distribution. The empirical function has negligible observed scatter. 
Such a radial acceleration relation (RAR) is a postulate of the 
empirical MOND model\citep{MIL83,FAM12,MIL16}, which does not invoke 
dark matter (DM).  It implies some very simple natural law.
\par Observed deviations from Newton/Einstein galactic gravitation 
have been modeled by distributed but unobserved dark matter. 
Typical DM halo models imply centripetal radial acceleration 
$a=a_N+\Da$, a function of radius in an assumed spherical 
galactic halo. DM $\Da$ is added to baryonic Newtonian $a_N$.
DM fits to galactic rotation (orbital velocity vs. circular orbit
radius) depend on model parameters central density $\rho_0$ and core
radius $r_0$ for a DM halo distribution.  Observed data imply that the
surface density product $\rho_0 r_0\simeq 100M_\odot pc^{-2}$ is nearly
constant for a large range of galaxies \citep{KAF04,GEN09,DON09}. 
\par It will be shown here that the conformal postulate for gravitation 
and the Higgs scalar field implies all of these regularities and 
derives values of the required parameters, without dark matter.

\section{Formalism for unified conformal theory}
\par Gravitational phenomena that cannot be explained by general
relativity as formulated by Einstein are attributed to cold dark
matter in the currently accepted $\Lambda$CDM paradigm for cosmology. 
Dark energy $\Lambda$ remains without an explanation.
The search for tangible dark matter has continued for many years with
no conclusive results\citep{SAN10}.
\par Consideration of an alternative paradigm is motivated by this
situation.  The postulate of universal conformal symmetry, requiring 
local Weyl scaling covariance\citep{WEY18} for all massless elementary 
physical fields\citep{NES13} is a falsifiable alternative.                                                                                                                                                                                                                      
Conformal symmetry, already valid for fermion and gauge boson fields
\citep{DEW64}, is extended to both the metric tensor field of
general relativity and the Higgs scalar field of elementary-particle
theory\citep{CAG98}.  This postulate is satisfied by conformal
gravity\citep{MAK89,MAN91,MAN06,MAN12} and by the conformal Higgs 
model\citep{NESM1,NES13}, eliminating the need for dark matter, with 
no novel elementary fields. The Higgs symmetry-breaking mechanism 
produces an effective cosmological constant driving Hubble expansion.
\par For fixed coordinates $x^\mu$, local Weyl scaling is defined by
$g_{\mu\nu}(x)\to g_{\mu\nu}(x)\Omega^2(x)$\citep{WEY18} for arbitrary
real differentiable $\Omega(x)$. Conformal symmetry is defined by
invariant action integral $I=\int d^4x\sqrt{-g}{\cal L}$.
For any Riemannian tensor $T(x)$, $T(x)\to \Omega^d(x)T(x)+{\cal
R}(x)$ defines weight $d[T]$ and residue ${\cal R}[T]$. For a scalar
field, $\Phi(x)\to\Phi(x)\Omega^{-1}(x)$, so that $d[\Phi]=-1$.
Conformal Lagrangian density ${\cal L}$ must have weight
$d[{\cal L}]=-4$ and residue ${\cal R}[{\cal L}]=0$ up
to a 4-divergence\citep{MAN06}.
\par Variational theory for fields in general relativity is a
straightforward generalization of classical field theory\citep{NES03}.
Given scalar Lagrangian density ${\cal L}=\sum_a{\cal L}_a$, action 
integral $I=\int d^4x \sqrt{-g} {\cal L}$ is required to be stationary 
for all differentiable field variations, subject to appropriate boundary
conditions.  $g$ here is the determinant of metric tensor $g_{\mu\nu}$.
Standard conservation laws follow from the variational principle.
\par Gravitational field equations are determined by metric
functional derivative
$X^{\mu\nu}= \frac{1}{\sqrt{-g}}\frac{\delta I}{\delta g_{\mu\nu}}$.
Any scalar ${\cal L}_a$ determines energy-momentum tensor
$\Theta_a^{\mu\nu}=-2X_a^{\mu\nu}$, evaluated for a solution of the
coupled field equations.  Generalized Einstein equation
$\sum_aX_a^{\mu\nu}=0$ is expressed as
$X_g^{\mu\nu}=\half\sum_{a\neq g}\Theta_a^{\mu\nu}$.  Hence summed
trace $\sum_ag_{\mu\nu}\Theta_a^{\mu\nu}$ vanishes for exact field
solutions.  
Given $\delta{\cal L}=x^{\mu\nu}\delta g_{\mu\nu}$,
metric functional derivative
$\frac{1}{\sqrt{-g}}\frac{\delta I}{\delta g_{\mu\nu}}$ is
$X^{\mu\nu}=x^{\mu\nu}+\half{\cal L}g^{\mu\nu}$,
evaluated for a solution of the field equations.
Tensor $\Theta^{\mu\nu}=-2X^{\mu\nu}$ is symmetric.
\revision{
\par Conformal gravity was introduced by Weyl\citep{WEY18,WEY18a} with 
the prospect of uniting electromagnetic theory and
gravitation.  This was shown to fail by Einstein, which side-tracked CG
for many years.  It was reconsidered by Mannheim as a possible source
of observed deviations from general relativity\citep{MAN06,MAN12}.
The formidable Lagrangian density, a quadratic contraction of the 
Weyl tensor, simplified by elimination of a 4-divergence\citep{LAN38}, 
reduces to two terms, quadratic respectively in the Ricci tensor and
Ricci scalar\citep{MAN06}: 
${\cal L}_g=-2\alpha_g( R^{\mu\nu}R_{\mu\nu}-\thrd R^2 )$.  Conformal
symmetry fixes the relative coefficient of the two quadratic terms.

\par In the conformal Higgs model, the unique Lagrangian density for
scalar field $\Phi$ adds a gravitational term\citep{MAN06,NESM1,NESM2} 
to Higgs $-V$\citep{CAG98}.  The conformal Lagrangian density is
$(\partial_\mu\Phi)^\dag\partial^\mu\Phi-\sxth R\Phi^\dag\Phi$,
where Ricci scalar $R=g_{\mu\nu}R^{\mu\nu}$ is the trace 
of gravitational Ricci tensor $R^{\mu\nu}$.
This is augmented by Higgs
$\Delta{\cal L}_\Phi=
(w^2-\lambda\Phi^\dag\Phi)\Phi^\dag\Phi$.
}   
\par Metric tensor $g_{\mu\nu}$ is determined by conformal field 
equations derived from ${\cal L}_g+{\cal L}_\Phi$\citep{NESM3},
driven by energy-momentum tensor $\Theta_m^{\mu\nu}$, where
subscript $m$ refers to conventional matter and radiation.
The gravitational field equation within halo radius $r_H$ is
\begin{eqnarray}
X_g^{\mu\nu}+X_\Phi^{\mu\nu}=\half\Theta_m^{\mu\nu}.
\end{eqnarray}
Defining mean galactic source density ${\bar\rho}_G$ and
residual density ${\hat\rho}_G=\rho_G-{\bar\rho}_G$
and assuming
$\Theta_m^{\mu\nu}(\rho)\simeq
 \Theta_m^{\mu\nu}({\bar\rho})+\Theta_m^{\mu\nu}({\hat\rho})$,
solutions for $r\leq r_G$ of the two equations
\begin{eqnarray} \label{Twoeqs}
X_\Phi^{\mu\nu}=\half\Theta_m^{\mu\nu}({\bar\rho}_G)
\nonumber\\
X_g^{\mu\nu}=\half\Theta_m^{\mu\nu}({\hat\rho}_G)
\end{eqnarray}
imply a solution of the full equation.
The $X_\Phi$ equation has an exact solution for source density
${\bar\rho}$ in uniform, isotropic geometry
(FLRW metric tensor)\citep{NESM1,NES13}.
The two field equations are made compatible by imposing appropriate 
boundary conditions.  Fitting constants of integration 
of source-free $X_g$ at $r_G$ \citep{MAK89}   
extends a combined solution out to halo radius $r_H$.
\par Conformal gravity is described in Schwarzschild metric\citep{MAK89}
\begin{eqnarray}
ds_{ES}^2=-B(r)c^2dt^2+\frac{dr^2}{B(r)}+r^2d\omega^2, 
\end{eqnarray}
which defines
Schwarzschild potential $B(r)$, assuming spherical symmetry.
The conformal Higgs model is described in FLRW metric
\begin{eqnarray}
ds_\Phi^2=-c^2dt^2+a^2(t)(\frac{dr^2}{1-kr^2}+r^2d\omega^2),
\end{eqnarray}
which defines Friedmann scaling factor $a(t)$.
\revision{
The Higgs model breaks both gauge and conformal symmetry.
}
Compatibility requires a hybrid metric with $k=0$  such as
\begin{eqnarray}
ds^2=-B(r)c^2dt^2+a^2(t)(\frac{dr^2}{B(r)}+r^2d\omega^2),
\end{eqnarray}

\revision{
\section{Depleted galactic halo model}
\par In the $\Lambda$CDM model, an isolated galaxy is considered to be
surrounded by a much larger spherical dark matter halo.  
What is actually observed is a halo of gravitational field that
deflects photons in gravitational lensing and increases the velocity
of orbiting mass particles. Conformal theory\citep{MAN06,MAN12,NES13}, 
which modifies both Einstein-Hilbert general relativity and the Higgs 
scalar field model, supports an alternative interpretation of lensing 
and anomalous rotation as gravitational effects due to depletion of the 
cosmic background by concentration of diffuse primordial mass into an 
observed galaxy\citep{NESM3}.
\par In standard theory, Poisson's equation determines a distribution
of dark matter for any unexplained gravitational field.  If dark 
matter interacts only through gravity, the concept of dark matter
provides a compact description of observed phenomena, not a falsifiable
explanation.  Postulated universal conformal symmetry\citep{NES13} 
promises a falsifiable alternative, free of novel elementary fields.
\par As a galaxy forms, background matter density $\rho_m$ condenses
into observed galactic density $\rho_g$.  Conservation of mass and
energy requires total galactic mass $M$ to be missing from a depleted
background.  Since the primordial density is uniform and isotropic, the
depleted background can be modeled by an empty sphere of radius $r_H$, 
such that $4\pi\rho_m r_H^3/3=M$.  In particular, the integral of 
$\rho_g-\rho_m$ must vanish.  The resulting gravitational effect, 
derived from conformal theory, describes a dark depleted halo.   
This halo model accounts for the otherwise remarkable fact that 
galaxies of all shapes are embedded in essentially spherical haloes. 
Conformal theory relates the acceleration parameter of anomalous 
galactic rotation to observed acceleration of Hubble expansion.
\par Assuming that galactic mass is concentrated within an average 
radius $r_g$, the ratio of radii $r_H/r_g$ should be large, the 
cube root of the mass-density ratio $\rho_g/\rho_m$.  Thus if the latter
ratio is $8^3=512$, a galaxy of radius 15kpc would be accompanied by a  
halo of radius $8\times 15=120$kpc.  Equivalence of galactic and
deleted halo mass resolves the paradox for $\Lambda$CDM that, despite 
any interaction other than gravity, the amount of dark matter inferred 
for a galactic halo is strongly correlated with the galactic luminosity 
or baryonic mass\citep{SAL07,MCG05a}.
}

\section{Conformal gravity}

\par Given spherical mass-energy density enclosed within $r\leq\R$,
the $X_g$ field equation in the ES metric is\citep{MAK89,MAN91}
\begin{eqnarray}
 \dar^4(rB(r))=rf(r), 
\end{eqnarray}
where $f(r)$ is determined by the source energy-momentum tensor.
\par An exact solution of the tensorial $X_g$ equation, for source-free 
$r\geq\R$, derived by\citep{MAK89,MAN91,MAN06}, is
\begin{eqnarray}
 y_0(r)=rB(r)=-2\beta+\alpha r+\gamma r^2-\kappa r^3.
\end{eqnarray}
\revision{
This adds two constants of integration to the 
classical external potential: nonclassical radial acceleration 
$\gamma$ and halo cutoff parameter $\kappa$\citep{MAN06,NESM3}.
}
Derivative functions $y_i(r) =\partial^i_r(rB(r)),0\leq i\leq3$
satisfy differential equations 
\begin{eqnarray}
 \dar y_i=y_{i+1},0\leq i\leq2, \nonumber\\
 \dar y_3=rf(r) .
\end{eqnarray}
Independent constants $a_i=y_i(0)$ determine coefficients
$\beta,\alpha,\gamma,\kappa$ such that at endpoint $\R$
\begin{eqnarray}
 y_0(\R)=-2\beta+\alpha \R+\gamma \R^2-\kappa \R^3,\nonumber\\
 y_1(\R)=\alpha+2\gamma \R-3\kappa \R^2,\nonumber\\
 y_2(\R)=2\gamma-6\kappa \R,\nonumber\\
 y_3(\R)=-6\kappa .
\end{eqnarray}
\par To avoid a singularity at the origin, $a_0=0$. 
$\gamma$ and $\kappa$ must be determined consistently with
the $X_\Phi$ equation.  Specified values can be fitted by adjusting
$a_1,a_2,a_3$, subject to $a_0=0,\alpha=1$\citep{MAN06}.
The solution for $B(r)$ remains regular if $\gamma/\beta\simeq 0$.
If Ricci scalar $R(r\to0)$ is unconstrained, a particular solution 
for $r\leq\R$ is given with specified constants 
$a_0=0,\alpha,\gamma,\kappa$\citep{NESM5}:
\begin{eqnarray}
 rB(r)=y_0(r)=\alpha r-\sxth\int_0^rq^4fdq-\half r\int_r^\R q^3fdq  
\nonumber\\
+\gamma r^2+\half r^2\int_r^\R q^2fdq
-\kappa r^3-\sxth r^3\int_r^\R qfdq .
\end{eqnarray}
Integrated parameter $2\beta=\sxth\int_0^\R q^4fdq$. 
This particular solution differs from\citep{MAK89,MAN06},
replacing $-\half r^2\int_0^rq^2fdq$ by
$\gamma r^2+\half r^2\int_r^\R q^2fdq$\citep{NESM5}.
Constant $\gamma$ is determined independently of the CG equation.
At the galactic center, $r\to0$ implies regular $B(r)$ but singular 
Ricci $R(r)$ for this solution. This is consistent with development
of a massive black hole, recently observed for many galaxies. 
\par For uniform density ${\bar\rho}$ the Weyl tensor vanishes, 
so that $X_g^{\mu\nu}\equiv 0$ for a uniform, isotropic cosmos. 
Schwarzschild potential, 
\begin{eqnarray} \label{Beq}
B(r)=-2\beta/r+\alpha+\gamma r-\kappa r^2, 
\end{eqnarray}
is exact for a source-free region, outside
a spherical mass-energy source density\citep{MAK89,MAN91},
in particular within an empty galactic halo.

\par For a test particle in a stable exterior 
circular orbit with velocity $v$ the centripetal acceleration is
$a=v^2(r)/r=\half B'(r)c^2$.
Newtonian $B(r)=1-2\beta/r$, where $\beta=GM/c^2$, so that
classical acceleration $a_N=\beta c^2/r^2=GM/r^2$.  CG adds 
nonclassical $\Da=\half c^2\gamma-c^2\kappa r$, a universal 
constant when halo cutoff 
$2\kappa r/\gamma\simeq r/r_H$ 
can be neglected \citep{NESM5}.  Orbital velocity
squared is the sum of $v^2(a_N;r)$ and $v^2(\Delta a;r)$, which
cross with equal and opposite slope at some $r=r_{TF}$,
if $2\kappa r/\gamma$ can be neglected.  This  defines 
a flat range of $v(r)$ centered at stationary point $r_{TF}$, without
constraining behavior at large $r$.
\par MOND\citep{MIL83,SAN10,FAM12} modifies the Newtonian force
law for acceleration below an empirical scale $a_0$.
Using $y=a_N/a_0$ as independent variable,
for assumed universal constant $a_0\simeq 10^{-10}m/s^2$,
MOND postulates an interpolation function $\nu(y)$ such that observed
radial acceleration $a=f(a_N)=a_N\nu(y)$.
A flat velocity range approached
asymptotically requires $a^2\to a_Na_0$ as $a_N\to0$.
For $a_N\ll a_0$, MOND $v^4=a^2r^2\to GMa_0$, the empirical
baryonic Tully-Fisher relation\citep{TAF77,MCG05,MCG11,OCM18}.
\par In conformal gravity (CG), centripetal acceleration
$a=v^2/r$ determines exterior orbital velocity
$v^2/c^2=ra/c^2=\beta/r+\half\gamma r-\kappa r^2$,
compared with asymptotic $ra_N/c^2=\beta/r$.
Assuming Newtonian constant $\beta=\beta_N$ and neglecting 
$2\kappa r/\gamma$, the slope of $v^2(r)$ vanishes at
$r^2_{TF}=2\beta/\gamma$.  This implies that
$v^4(r_{TF})/c^4=(\beta/r_{TF}+\half\gamma r_{TF})^2=
2\beta\gamma$ \citep{MAN97,OCM18}.
This is the Tully-Fisher relation, exact at stationary point
$r_{TF}$ of the $v(r)$ function.   The implied MOND acceleration
constant is $a_0=2\gamma c^2=1.14\times 10^{-10} m/s^2$\citep{NESM5}.

\section{The conformal Higgs model}
\par The Newton-Einstein model of galactic growth and interaction is 
substantially altered by conformal theory\citep{NESM1,NESM3}.
A gravitational halo is treated as a natural consequence of
galaxy formation.  Condensation of matter from the primordial uniform
mass-energy distribution leaves a depleted sphere that has an explicit
gravitational effect. The implied subtracted mass, which integrates to 
the total galactic mass, cannot be ignored.  If this mass were 
simply removed, the analogy to vacancy scattering in solids implies a 
lensing effect.  A background geodesic is no longer a geodesic 
in the empty sphere\citep{NESM3}.
\par The conformal gravitational field equation is 
\begin{eqnarray} \label{cfeq}
X_g^{\mu\nu}+X_\Phi^{\mu\nu}=\half\Theta_m^{\mu\nu},
\end{eqnarray}
where index m refers to matter and radiation.   
An exact solution inside the depleted halo radius is given by
\begin{eqnarray} \label{geq}
X_g^{\mu\nu}=\half\Theta_m^{\mu\nu}(\rho_g),
X_\Phi^{\mu\nu}=0, r\leq r_H.
\end{eqnarray}
The exact source-free solution\citep{MAK89} of the $X_g$ equation
is valid in the external halo, $r_g\leq r\leq r_H$, because $\rho_g$ 
vanishes.  Constants of integration fitted at $r_g$ and $r_H$ 
determine radial acceleration in this external halo.
\par In uniform, isotropic geometry with uniform mass/energy density
${\bar\rho}$, ${\cal L}_\Phi$
implies a modified Friedmann equation 
\citep{FRI22,NESM1,NES13} for cosmic distance scale factor $a(t)$,
in FLRW metric
\begin{eqnarray}
ds_\Phi^2=-c^2dt^2+a^2(t)(\frac{dr^2}{1-kr^2}+r^2d\omega^2),
\end{eqnarray}
which determines Friedmann scaling factor $a(t)$.
With $a(t_0)=1$ at present time $t_0$:
\begin{eqnarray} \label{MFeq}
 \frac{{\dot a}^2}{a^2}+\frac{k}{a^2}-\frac{\ddot a}{a}
    =\frac{2}{3}({\bar\Lambda}+{\bar\tau}c^2{\bar\rho}(t)).
\end{eqnarray} 
Consistency with the $X_g$ equation requires $k=0$.
${\bar\Lambda}=\frac{3}{2}w^2\geq 0$ and 
${\bar\tau}\sim-3/\phi_0^2\leq 0$ are determined by 
parameters of the Higgs model\citep{NESM1}. For uniform $\rho=0$ 
within the halo radius this solves the second of Eqs.(\ref{geq}).
\par Dividing the modified equation by $({\dot a}/a)^2$ determines 
dimensionless sum rule $\Omega_m+\Omega_\Lambda+\Omega_k+\Omega_q=1$.
Scale factor $a(t)$ determines dimensionless Friedmann acceleration 
weight $\Omega_q(t)=\frac{{\ddot a}a}{{\dot a}^2}$\citep{NESM1}. 
Radiation energy density is included in $\Omega_m$ here.
The dimensionless weights are $\Omega_m(t)$ for mass density, 
$\Omega_k(t)$ for curvature, and $\Omega_\Lambda(t)$ for dark energy, 
augmented by acceleration weight $\Omega_q(t)$.
Galactic rotation velocities, observed at relatively small redshifts,
determine radial acceleration values which can be compared with 
acceleration weights inferred from Hubble expansion data.
Omitting $\Omega_m$ completely, with $k=0$, conformal sum rule 
$\Omega_\Lambda(t)+\Omega_q(t)=1$ fits observed data accurately for 
redshifts $z\leq 1$ (7.33Gyr) \citep{NESM1,NESM2,NESM5}.
This eliminates any need for dark matter to explain Hubble expansion.   
\par The $X_\Phi$ equation includes dark energy, present regardless
of density $\rho$.  This produces centrifugal acceleration
of background Hubble expansion\citep{NESM1,NES13}, which must be 
subtracted off in order to compute observable radial acceleration.
Observed geodesics, whose bending determines the extragalactic
centripetal acceleration responsible for lensing and anomalous orbital 
rotation velocities, are defined relative to the cosmic background.
\par Defining $\frac{{\dot a}}{a}=h(t)H_0$, 
for Hubble constant $H_0$, it is convenient to use 
Hubble units, such that $c=\hbar=1$, and dimensionless
$a(t_0)=1$, $h(t_0)=1$. The units for frequency, energy, and 
acceleration are $H_0,\hbar H_0,cH_0$, respectively.
Setting $\Omega_k=\Omega_m=0$, for 
$\alpha=w^2=\Omega_\Lambda(t_0)$, the CHM Friedmann 
equation\citep{NESM1,NES13} is
\begin{eqnarray} \label{cFeq}
 \frac{{\dot a}^2}{a^2}-\frac{\ddot a}{a}=
    \frac{2}{3}{\bar\Lambda}=\alpha.
\end{eqnarray}
\par In Hubble units, Eq.(\ref{cFeq}) reduces to 
$\frac{d}{dt}h(t)=-\alpha$.
The explicit solution for $t\leq t_0$ is
\begin{eqnarray}
h(t)=\frac{{\dot a}}{a}(t)=\frac{d}{dt}\ln a(t)=1+\alpha(t_0-t),
\nonumber\\
\ln a(t)=-(t_0-t)-\half\alpha(t_0-t)^2,
\nonumber\\
a(t)=\exp[-(t_0-t)-\half\alpha(t_0-t)^2].
\end{eqnarray}
\par From the definition of redshift $z(t)$: \\
$1+z(t)=\frac{1}{a(t)}=\exp[(t_0-t)+\half\alpha(t_0-t)^2],\\
(t_0-t)^2+\frac{2}{\alpha}(t_0-t)=\frac{2}{\alpha}\ln(1+z)$.
\par In Hubble units with $\frac{{\dot a}}{a}(t_0)=1$, this implies
\begin{eqnarray}   \label{dteq}
t_0-t(z)=(\sqrt{2\alpha\ln(1+z)+1}-1)/\alpha,
\nonumber\\
\frac{dt}{dz}=\frac{-1}{(1+z)\sqrt{2\alpha\ln(1+z)+1}}.
\end{eqnarray}
If $\alpha=0.732$ and $z=1$, $(t_0-t)/H_0=7.33Gyr$, for
$H_0=67.8\pm 0.9 km/s/Mpc=2.197\times 10^{-18}/s$\citep{PLC15}.
\par Neglecting both curvature weight $\Omega_k$ and cosmic 
mass/energy weight $\Omega_m$, conformal sum rule
$\Omega_\Lambda+\Omega_q=1$ holds for acceleration weight
$\Omega_q=\frac{{\ddot a}a}{{\dot a}^2}$. For $\Omega_k=0$, 
luminosity distance $d_L(z)=(1+z)\chi(z)$\citep{NESM2}, where
\begin{eqnarray} \label{Chi_z}
\chi(z)=
\int_{t(z)}^{t_0}\frac{dt}{a(t)}=
 \int_z^0 d\z(1+\z)\frac{dt}{dz}(\z)=
\nonumber\\
 \int_0^z\frac{d\z}{\sqrt{2\alpha\ln(1+\z)+1}}.
\end{eqnarray}
\par Evaluated for $\alpha=\Omega_\Lambda(t_0)=0.732$,
the fit to scaled luminosity distances $H_0d_L/c$ from Hubble
expansion data\citep{MAN03,NESM1} is given in Table \ref{TabA1}. 
\begin{table}[h]
\caption{Scaled luminosity distance fit to Hubble data} \label{TabA1}
\begin{tabular}{lcccc}
 &                 &          &Theory     &Observed    \\
z& $\Omega_\Lambda$&$\Omega_q$&$H_0d_L/c$Eq.(\ref{Chi_z})
 &$H_0d_L/c$\citep{MAN03}\\ \hline
0.0& 0.732& 0.268& 0.0000& 0.0000\\
0.2& 0.578& 0.422& 0.2254& 0.2265\\
0.4& 0.490& 0.510& 0.5013& 0.5039\\
0.6& 0.434& 0.566& 0.8267& 0.8297\\
0.8& 0.393& 0.607& 1.2003& 1.2026\\
1.0& 0.363& 0.637& 1.6209& 1.6216
\end{tabular}
\end{table}
Observed redshifts have been fitted to an analytic 
function\citep{MAN03} with
statistical accuracy comparable to the best standard $\Lambda$CDM fit,
with $\Omega_m=0$. Table(\ref{TabA1}) compares CHM $d_L(z)$ to this
Mannheim function.  Because ${\bar\tau}$ is negative\citep{MAN06,NESM1},
cosmic acceleration $\Omega_q$ remains positive (centrifugal) back to
the earliest time\citep{NESM1}.
\par Conformal Friedmann Eq.({\ref{MFeq}})\citep{NESM1,NESM2} determines 
cosmic acceleration weight $\Omega_q$. With both weight parameters 
$\Omega_k$ and $\Omega_m$ set to zero, Eq.(\ref{MFeq}) fits scaled 
Hubble function $h(t)=H(t)/H_0$ for redshifts $z\leq 1$,
Table(\ref{TabA1}), as accurately as standard $\Lambda$CDM, with only one 
free constant.  This determines Friedmann weights, at present time 
$t_0$, $\Omega_\Lambda=0.732, \Omega_q=0.268$ \citep{NESM1}.
Hubble constant $H(t_0)=H_0=2.197\times 10^{-18}/s$ \citep{PLC15}
is independent of these data.
For uniform primordial energy-momentum source density $\rho_m$,
the dimensionless sum rule\citep{NESM1} with $\Omega_k=0$ determines
$\Omega_q(\rho_m)=1-\Omega_\Lambda-\Omega_m$ in the cosmic background,  
and $\Omega_q(0)=1-\Omega_\Lambda$ in the depleted halo\citep{NESM3}.                                     
\par Schwarzschild $B(r)$ parameter $\gamma>0$, independent of galactic 
mass and structure, implies centripetal acceleration due to an isotropic 
cosmological source\citep{MAN97}. The parametrized gravitational field 
forms a spherical halo\citep{NES13}. The depleted halo model removes a 
particular conceptual problem in fitting $B(r)$ parameters 
$\gamma,\kappa$ to galactic rotation data\citep{MAN97,MAN06,MAO11}.  
In empirical parameter $\gamma=\gamma^*N^*+\gamma_0$, $\gamma_0$ does 
not depend on galactic mass, so must be due to the surrounding 
cosmos\citep{MAN97}.  Since the interior term, coefficient $\gamma^*N^*$, 
is centripetal, one might expect the term in $\gamma_0$ to be 
centrifugal, describing attraction to exterior sources.  However, 
coefficient $\gamma$, derived here for a depleted halo, determines net 
centripetal acceleration, in agreement with observation\citep{NESM5}. 
Recent galactic rotation results for galaxies with independently 
measured mass\citep{MLS16} imply that $\gamma$ cannot depend on galactic
mass\citep{NESM7}.
\revision{
\par The conformal Friedmann cosmic evolution
equation implies dimensionless cosmic acceleration parameters 
$\Omega_q(\rho)$\citep{NESM3} which are locally constant but differ 
across the halo boundary $r_H$.  Smooth evolution of the cosmos implies 
observable particle acceleration $\gamma$ within $r_H$ proportional to 
$\Omega_q(in)-\Omega_q(out)=\Omega_q(0)-\Omega_q(\rho_m)$.  Uniform 
cosmological $\rho_m$ implies constant $\gamma$ for $r\leq r_H$,
independent of galactic mass \citep{NESM5}.  This result is 
consistent with recent observations of galactic rotational velocities
for galaxies with directly measured mass\citep{MLS16,NESM7}. 
\par From the CHM, observed nonclassical gravitational acceleration 
$\half\gamma c^2$ in the halo is proportional to 
$\Delta\Omega_q=\Omega_q(0)-\Omega_q(\rho_m)=
\Omega_m(\rho_m)$\citep{NESM3}, where, given $\rho_m$ and $H_0$, 
$\Omega_m(\rho_m)=
\frac{2}{3}\frac{{\bar\tau}c^2\rho_m}{H^2_0}$\citep{NESM1}.
Positive $\rho_m$ implies $\Omega_m<0$ because coefficient 
${\bar\tau}<0$ \citep{MAN06,NESM1}.  
Thus the depleted halo model determines constant $\gamma$ from
uniform universal cosmic baryonic mass density $\rho_m/c^2$,
which includes radiation energy density here. 
\par $\Omega_m<0$ implies centripetal acceleration, converted from 
Hubble units, $\half\gamma c^2=-cH_0\Omega_m(\rho_m)$\citep{NESM3}.
Hence $\Delta\Omega_q=\Omega_m<0$ is consistent with nonclassical
centripetal acceleration, confirmed by inward
deflection of photon geodesics observed in gravitational
lensing \citep{NESM3}.  This logic is equivalent to requiring 
radial acceleration to be continuous across halo boundary $r_H$:
\begin{eqnarray}
\half\gamma c^2-cH_0\Omega_q(0)=-cH_0\Omega_q(\rho_m).
\end{eqnarray}
Signs here follow from the definition of $\Omega_q$ as centrifugal
acceleration weight.
}
\par For a single spherical solar mass isolated in a galactic halo,
mean internal mass density ${\bar\rho}_\odot$ within $r_\odot$
determines an exact solution of the conformal Higgs gravitational
equation, giving internal acceleration 
$\Omega_q({\bar\rho}_\odot)$.
Given $\gamma$ outside $r_\odot$,
continuous acceleration across boundary $r_\odot$,
\begin{eqnarray}  
\half\gamma_{\odot,in}c^2-cH_0\Omega_q({\bar\rho}_\odot)=
 \half\gamma c^2-cH_0\Omega_q(0),
\end{eqnarray}
determines constant $\gamma_{\odot,in}$ valid inside $r_\odot$.
$\gamma_{\odot,in}$ is determined by local mean source density
${\bar\rho}_\odot$. $\gamma$ in the halo is not changed.  Its value is 
a constant of integration that cannot vary in the source-free 
halo\citep{NESM3,NESM5}.  Hence there is no way to determine
a mass-dependent increment to $\gamma$.  This replaces the usually 
assumed $\gamma_0+N^*\gamma^*$ by $\gamma_H$ 
determined at halo boundary $r_H$.

\section{Values of relevant parameters}
\revision{
\par Anomalous rotation velocities for 138 galaxies are fitted using
only four universal CG parameters $\beta^*,\gamma^*,\gamma_0,\kappa$
\citep{MAN06,MAN12,MAO11,MAO12,OAM15} such that
$\beta=N^*\beta^*=GM/c^2,\gamma=\gamma_0+N^*\gamma^*$.
$N^*$ is galactic baryonic mass $M$ in solar mass units.
Inferred parameter values \citep{MAN06,MAO12}, 
\begin{eqnarray}
\beta^*=1.475\times 10^3 m,
\gamma_0=3.06\times 10^{-28}/m, \nonumber\\
\gamma^*=5.42\times 10^{-39}/m,
\kappa  =9.54\times 10^{-50}/m^2,
\end{eqnarray}
fit conformal gravity to galactic rotation velocities. The observed 
radial acceleration relation (RAR)\cite{MLS16} is consistent with the 
CHM conclusion that $\gamma^*$ must vanish.  Then $\gamma$ is a 
universal constant\citep{NESM7}.  Using data for the Milky Way galaxy,
$\gamma=\gamma_0+N^*\gamma^*=6.35\times 10^{-28}/m$\citep{OAM15,MCG08}. 
\par $\zeta>0$ for computed $R(t)$\citep{NESM1} implies $\lambda<0$.
$\hbar\phi(t_0)=174GeV$\citep{AMS08}$=1.203\times 10^{44}\hbar H_0$ 
in Hubble units. For $\Omega_m=0$, $\zeta(t_0)=2\Omega_q(t_0)=0.536$. 
For $\lambda(t)=\zeta/(-2\phi^2)$ and $\phi(t_0)=\phi_0$,
dimensionless $\lambda(t_0)=-0.185\times 10^{-88}$.
}
\par It is widely assumed that negative Higgs $\lambda$ would imply 
an unstable physical vacuum, but the present analysis does not support
this conclusion. The conformal Higgs scalar field does not have a 
well-defined mass, instead inducing dynamical $w^2$, which acts as a 
cosmological constant determining dark energy.
Nearly constant ${\dot\phi}/\phi$ for redshifts $z\leq 1$ is shown here
to imply Higgs $\lambda<0$, consistent with a dynamical origin.  
Negative empirical $\lambda$ implies finite but `tachyonic'
mass for a scalar field fluctuation.  This does not support the
conventional concept of a massive Higgs particle\citep{NESM4}.
\par Scalar field $ZZ=g_{\mu\nu}Z^{\mu*}Z^\nu$  
interacts with neutral scalar field $WW=g_{\mu\nu}W_-^\mu W_+^\nu$ 
through exchange of quarks and leptons. There is no contradiction in 
treating resulting scalar diboson $W_2$ as an independent field or 
particle, in analogy to atoms, molecules, and nuclei.  Neutral $W_2$ 
dresses scalar field $\Phi$ to produce biquadratic Lagrangian term 
$\lambda(\Phi^\dag\Phi)^2$\citep{NESM4}. If the mass of $W_2$ is 
$125GeV$ the conformal Higgs $\lambda$ assumes its empirical value. 
This implies an alternative identification of the recently observed              
LHC resonance \citep{DJV12,ATL12,CMS12}.
The interacting bare fields produce relatively stable                                                                                                        
$W_2=WW\cos\theta_x+ZZ\sin\theta_x$ and complementary resonance
$Z_2=-WW\sin\theta_x+ZZ\cos\theta_x$\citep{NESM4}. $W_2$ can dress the 
bare $\Phi$ field while maintaining charge neutrality.  Assumed $W_2$
mass 125GeV implies $\lambda=-0.455\times 10^{-88}$\citep{NESM4},                        
consistent with empirical value $\lambda(t_0)\simeq-10^{-88}$.
 
\par
\revision{  
Parameters $w^2$ and $\lambda$ are defined
as positive constants in Higgs\citep{HIG64,CAG98}
\begin{eqnarray}\label{DeltaL_par}
\Delta{\cal L}_\Phi=-V(\Phi^\dag\Phi)=
w^2\Phi^\dag\Phi-\lambda(\Phi^\dag\Phi)^2.
\end{eqnarray}
Stationary action implies constant 
finite $\Phi^\dag\Phi=\phi_0^2=w^2/2\lambda$, where
$\hbar\Phi$ and $\hbar w$ are energies.
The conformal Higgs model\citep{NESM1} acquires an additional 
term in $\Delta{\cal L}_\Phi$:
$-\sxth R \Phi^\dag\Phi$\citep{MAN06} for gravitational Ricci scalar 
$R=g_{\mu\nu}R^{\mu\nu}$. The modified Lagrangian, defined for 
neutral gauge field $Z_\mu$\citep{NESM2},
determines $w^2$, which becomes a cosmological constant
in the conformal Friedmann cosmic evolution equation\citep{NESM1}. 
Lagrangian term $w^2\Phi^\dag\Phi$ is due to induced neutral 
$Z_\mu$ field amplitude\citep{NESM2}, which dresses the bare 
scalar field. Finite $w^2$ breaks conformal symmetry, but does not 
determine a Higgs mass.
}
\par The conformal scalar field equation
including parametrized $\Delta{\cal L}_\Phi$ is\citep{MAN06,NESM1}
\begin{eqnarray}\label{Phieq}
\frac{1}{\sqrt{-g}}\partial_\mu(\sqrt{-g}\partial^\mu\Phi)=
 (-\sxth R+w^2-2\lambda\Phi^\dag\Phi)\Phi.
\end{eqnarray}
Ricci scalar $R$ introduces gravitational effects.
Only real-valued solution $\phi(t)$ is relevant in uniform,
isotropic geometry. Defining $V(\phi)=
 \half(\zeta+\lambda\phi^2)\phi^2$,
where $\zeta(t)=\sxth R-w^2$, the field equation is
\begin{eqnarray} \label{phieq}
\frac{{\ddot\phi}}{\phi}
+3\frac{{\dot a}}{a}\frac{{\dot\phi}}{\phi}=
 -(\zeta(t)+2\lambda\phi^2).
\end{eqnarray}
\revision{
\par Omitting $R$ and assuming constant $\lambda>0$ and $w^2$, Higgs
solution $\phi_0^2=w^2/2\lambda$\citep{HIG64} is exact.
All time derivatives drop out.  In the conformal scalar field equation,
cosmological time dependence of Ricci scalar $R(t)$, determined by the 
CHM Friedmann cosmic evolution equation, introduces nonvanishing time 
derivatives and implies $\lambda<0$\citep{NESM1}.  An exact solution 
for $\phi(t)$ should include the time derivative terms in $\zeta(t)$.  
Neglecting them here limits the argument to redshift $z\leq 1$. 
}
\par
\revision{
Ricci scalar $R(t)$ drops out  of the scalar field equations if time 
derivatives are neglected.  This justifies an estimate of Higgs
parameter $w^2$\citep{NESM1}.  Dressing scalar field $\Phi$
by neutral gauge field $Z^\mu$ produces Lagrangian term 
$w^2\Phi^\dag\Phi$.
} 
Defining $\Delta I_\Phi=\int d^4x\sqrt{-g}\Delta{\cal L}_\Phi$
from Eq.(\ref{DeltaL_par}), the parametrized
effective potential term in the  scalar field equation is given by
\begin{eqnarray} \label{DPhieq}
\frac{1}{\sqrt{-g}}\frac{\delta\Delta I_\Phi}{\delta\Phi^\dag}=
 (w^2-2\lambda\Phi^\dag\Phi)\Phi . 
\end{eqnarray}
Gauge invariance replaces bare derivative $\partial_\mu$
by gauge covariant derivative\citep{CAG98}
\begin{eqnarray}
D_\mu=\partial_\mu-\frac{i}{2}g_z Z_\mu.
\end{eqnarray}
This retains ${\cal L}_Z$ in terms of
$Z_{\mu\nu}=\partial_\mu Z_\nu-\partial_\nu Z_\mu$ 
and augments conformal 
\begin{eqnarray}
{\cal L}^0_\Phi=
(\partial_\mu\Phi)^\dag \partial^\mu\Phi-\sxth R\Phi^\dag\Phi 
\end{eqnarray} 
by coupling term 
\begin{eqnarray} \label{DelD}
\Delta{\cal L}=
(D_\mu\Phi)^\dag D^\mu\Phi -(\partial_\mu\Phi)^\dag \partial^\mu\Phi=
\nonumber\\
 \frac{i}{2}g_z Z_\mu^*\Phi^\dag\partial^\mu\Phi
-\frac{i}{2}g_z Z^\mu(\partial_\mu\Phi)^\dag\Phi
+\frth g_z^2\Phi^\dag Z_\mu^* Z^\mu\Phi.
\end{eqnarray}
\par Parametrized for a generic complex vector field\citep{CAG98}, 
\begin{eqnarray} \label{DelZ}
\Delta{\cal L}_Z=
 \half m^2_Z Z^*_\mu Z^\mu-\half(Z^*_\mu J_Z^\mu+Z^\mu J^*_{Z\mu}),
\end{eqnarray}
given mass parameter $m_Z$ and source current density $J_Z^\mu$.
The field equation for parametrized $Z^\mu$ is\citep{CAG98}
\begin{eqnarray}\label{Zeq}
 \partial_\nu Z^{\mu\nu}=
\frac{2}{\sqrt{-g}}\frac{\delta\Delta I_Z}{\delta Z_\mu^*}=
m_Z^2 Z^\mu-J_Z^\mu . 
\end{eqnarray} 
$\Delta{\cal L}$ determines parameters for field $Z^\mu$:
\begin{eqnarray}\label{Zcdeq}
 \frac{2}{\sqrt{-g}}\frac{\delta\Delta I}{\delta Z_\mu^*}=
 \half g_z^2\Phi^\dag\Phi Z^\mu+ig_z\Phi^\dag\partial^\mu\Phi. 
\end{eqnarray}
This implies Higgs mass formula $m_Z^2=\half g_z^2\Phi^\dag\Phi$
and field source density
$ J_Z^\mu=-ig_z\Phi^\dag\partial^\mu\Phi$.
Using $\Delta{\cal L}$ derived from the covariant derivative,
\begin{eqnarray}
 \frac{1}{\sqrt{-g}}\frac{\delta\Delta I_\Phi}{\delta \Phi^\dag}=
\frth g_z^2Z_\mu^*Z^\mu \Phi
+\frac{i}{2}g_z(Z_\mu^*+Z_\mu)\partial^\mu\Phi.
\end{eqnarray}
Comparison with Eq.(\ref{DPhieq}) implies 
$w^2=\frth g_z^2 Z_\mu^*Z^\mu$.
Neglecting derivatives of $Z^\mu$,
Eq.(\ref{Zeq}) reduces to
\begin{eqnarray} 
Z^\mu= J_Z^\mu/m_Z^2 =
-\frac{2i}{g_z}\Phi^\dag\partial^\mu\Phi/\Phi^\dag\Phi.
\end{eqnarray}
Then $|Z^0|^2=\frac{4}{g_z^2}(\frac{{\dot\phi}}{\phi})^2$,
so that the scalar field equation implies nonvanishing 
$w^2=\frth g_z^2|Z^0|^2=(\frac{{\dot\phi}}{\phi})^2$.
Implied pure imaginary $Z^\mu$ does not affect parameter $\lambda$.

\section{Alternative models of observed data}
\par Numerous dark matter studies of galactic halo gravitation depend on 
models with  core radius $r_0$ and central density $\rho_0$.  Central
surface density product $\rho_0 r_0$ is found to be nearly a universal 
constant of order $100M_\odot pc^{-2}$ for a large range of galaxies   
\citep{KAF04,GEN09,DON09}.
\revision{  
Conformal theory implies nonclassical centripetal acceleration $\Da$,
for Newtonian $a_N$ due to observable baryonic matter.  Neglecting 
the halo cutoff for $r\ll r_H$, conformal $\Da$ is 
constant over the  halo and $a=a_N+\Da$ is a universal function, 
consistent with a recent study of galaxies with independently measured 
mass\citep{MLS16}. An equivalent dark matter source would be a pure cusp 
distribution with cutoff parameter determined by halo boundary radius.
This is shown here to imply universal central surface density for any 
dark matter core model.
\par CG implies $\Da=\half\gamma c^2(1-r/r_H)$ in a depleted halo.
For $r\ll r_H$,
CG acceleration constant $\gamma$ \citep{MAN06,NESM7} predicts
$\Da=\half\gamma c^2$ \citep{NESM3,NESM7} for all DM models.

\par Uniform constant $\Da$ puts a severe constraint  
on any DM model.  The source density must be of the form
$\xi/r$, a pure radial cusp \citep{NESM7,NESM5}, where constant
$\xi={\Da}/{2\pi G}=0.06797kg/m^2=32.535M_\odot/pc^2$ is
determined by CG parameter $\gamma$, with CODATA 
$G=6.67384\times 10^{-11}m^3s^{-2}kg^{-1}$ \citep{COD12}.
\par A DM galactic model equivalent to conformal theory would imply 
uniform DM radial acceleration $\Da=2\pi G\xi$, attributed to 
radial DM density $\xi/r$ for universal constant $\xi$, modified at 
large galactic radius by a halo cutoff function.  Enclosed mass
$M_r=2\pi \xi r^2$ implies $r\Da/c^2=GM_r/r$. DM models avoid a 
distribution cusp by assuming finite central core density.

\par The distribution of source matter 
within a sphere cannot affect asymptotic gravitational acceleration.
For arbitrary $r_0$, asymptotic radial acceleration
is unchanged if mass within $r_0$ is redistributed to uniform 
density $\rho(r)$ within a sphere of this radius.  
Conformal density $\xi/r$ implies mass $M_0=2\pi \xi r_0^2$ in 
volume $V_0=\frac{4\pi}{3} r_0^3$. For a DM spherical model core 
that replaces a central cusp density, conformal theory implies constant 
$\rho(r_0)r_0=r_0M_0/V_0=3\xi/2$.
For assumed PI core DM density \citep{KAF04}
$\rho(r)=\rho_0 r_0^2/(r^2+r_0^2)$ ,
central $\rho_0=2\rho(r_0)$.  Hence for a PI core, 
$\rho_0r_0=3\xi=\frac{3\Da}{2\pi G}
 =0.204\times 10^{-2}kg/m^2= 97.6M_\odot pc^{-2}$,
independent of $r_0$.
}
                        
\par A radial acceleration relation RAR\citep{MLS16} 
is postulated by MOND\citep{MIL83}. Paradigms $\Lambda$CDM                                                                                                             
and CG/CHM both represent galactic radial acceleration by $a_N+\Da$,  
for baryonic Newtonian $a_N$. The RAR requires $\Da$  
to be a universal constant\citep{NESM7}, for $r\ll r_H$.  Given $a_N$, 
the three paradigms differ only by the behavior of
$\Da$ for large galactic radius. 
\par Milky Way Tully-Fisher radius $r_{TF}=17.2 kpc$. Halo radius
$r_H=107.8kpc$\citep{NESM3,NESM5}, for $r_G\simeq 15.0 kpc$.
Implied MOND constant $a_0=2\gamma c^2=1.14\times 10^{-10} m/s^2$.
Outside $r_G$, $a_N\simeq\beta c^2/r^2$.
Then a(CDM)$=a_N+\half\gamma c^2$, 
using empirical CG $\Delta a$ but omitting halo cutoff;
a(CG)$=a_N+\half\gamma c^2(1-r/r_H)$, including halo cutoff; and
a(MOND\citep{MLS16})$\simeq
 a_N/(1-e^{-\sqrt{a_N/a_0}})$,
just MOND with a particular interpolation function 
and $a_0=1.20\times 10^{-10}m/s^2$.
$a(CDM)$ is generic for any model with universal constant $\Delta a$.
\begin{table} [h]
\caption{Milky Way: radial acceleration a ($10^{-10}m/s^2$)}
 \label{Tab02}
\begin{tabular}{lccccccc}
r  &     &CDM&&CG&&MOND&\\
kpc&$a_N$& a     &$10^3\frac{v}{c}$
         & a     &$10^3\frac{v}{c}$
         & a     &$10^3\frac{v}{c}$\\
\hline
15& 0.376& 0.661& 0.584& 0.621& 0.566& 0.877& 0.672\\
20& 0.212& 0.497& 0.584& 0.444& 0.552& 0.617& 0.650\\
25& 0.135& 0.420& 0.601& 0.354& 0.551& 0.475& 0.638\\
30& 0.094& 0.379& 0.625& 0.300& 0.556& 0.385& 0.630\\
35& 0.069& 0.354& 0.652& 0.261& 0.560& 0.324& 0.624\\
40& 0.053& 0.338& 0.682& 0.232& 0.565& 0.279& 0.619\\
45& 0.042& 0.327& 0.717& 0.208& 0.567& 0.246& 0.616\\
50& 0.034& 0.319& 0.740& 0.187& 0.566& 0.219& 0.613
\end{tabular}
\end{table}
\par Table \ref{Tab02}\citep{NESM7} 
compares detailed predictions for the implied
external orbital velocity curve of the Milky Way
galaxy.  The CDM curve rises gradually, the CG curve remains 
remarkably flat, while the MOND\citep{MLS16} curve falls gradually
toward asymptotic velocity.

\section{Conclusions}
\revision{
The postulate of universal conformal symmetry is shown to be in good 
agreement with gravitational phenomena wherever tested. 
Implied values of Higgs scalar field parameters are 
also found to be valid.  The resulting gravitational paradigm can be 
considered a valid alternative to $\Lambda$CDM and MOND, justifying
wider study and application. Selection among theoretical models depends 
on improvement of the precision of observations of galactic rotation at 
large galactic radius, currently scarce and of limited accuracy.
}  



\vfill\eject
\end{document}